\documentclass[twocolumn,twocolappendix]{aastex631}

\usepackage{amsmath}

\usepackage{orcidlink}
\newcommand{\AuthorORCID}[2]{%
  \href{https://orcid.org/#2}{\mbox{#1~{\Large\orcidlink{#2}}\kern-0.5em}}%
  }
\usepackage{todonotes}
\presetkeys{todonotes}{inline}{}

\begin{document}

\title{Supernovae Exploding within Dense Extended Material: Early Emission Regimes and  Degeneracies in Parameter Inference from Observations}

\shorttitle{Early Emission of Supernovae in Extended Material}
\shortauthors{Wasserman \& Waxman}

\author{\AuthorORCID{Tal Wasserman}{0009-0005-7414-3965}}
\affiliation{Department of Particle Physics \& Astrophysics, Weizmann Institute of Science, Rehovot 76100, Israel}
\author{\AuthorORCID{Eli Waxman}{0000-0002-9038-5877}}
\affiliation{Department of Particle Physics \& Astrophysics, Weizmann Institute of Science, Rehovot 76100, Israel}

\begin{abstract}
Early light curves of many core-collapse supernovae (SNe) are thought to be powered by the interaction of the shock wave with optically thick extended material, either a bound envelope or preexplosion ejected circumstellar matter (CSM). We analytically analyze the early emission produced by a shock with velocity $v$ traversing a material of mass $M_\mathrm{e}$ and opacity $\kappa$ extending to radius $R_\mathrm{e}$, and show the emission varies qualitatively with varying $\tau_\mathrm{e}=\kappa\!M_\mathrm{e}/(4\pi\!R_\mathrm{e}^2)$: For $\tau_\mathrm{e}\gg\!c/v$ a shock breakout occurs near $R_\mathrm{e}$ producing an ``edge breakout"- a UV-dominated breakout burst followed by ``cooling emission" of the shock-heated material; for $\tau_\mathrm{e}\lesssim\!c/v$ a ``wind breakout" occurs- the breakout pulse is prolonged and followed by extended emission shifting from UV to X-ray as the shock becomes collisionless. We derive the dependence on $\{v,\kappa,M_\mathrm{e},R_\mathrm{e}\}$ of the duration and luminosity characterizing the different emission phases, and show that current observations typically do not allow inference of all parameters. In particular, since the optical bands lie in the Rayleigh-Jeans tail of radiation emitted during the cooling phase, the observed cooling luminosity depends weakly on $R_\mathrm{e}$, $\propto\!R_\mathrm{e}^{1/4}$, leading to $1-2$ orders of magnitude uncertainty in its inferred value. This suggests, e.g., that the common day-scale light curve features in Stripped-Envelope SNe do not necessarily imply material extending to $R_\mathrm{e}\sim10^3\!R_\odot$ and are often consistent with low-mass $R_\mathrm{e}\sim\!10^2\!R_\odot$ bound envelopes. Early multiband coverage (especially in UV/X-ray) can break these degeneracies; the forthcoming \emph{ULTRASAT} UV mission will allow inferring the properties of extended material around the population of SN progenitors.
\end{abstract}

\section{Introduction}\label{sec:intro}

The early, hours-to-day, supernovae (SNe) emission holds valuable information on the structure of the progenitors and their environment. If the progenitor star's envelope is characterized by a sharp stellar surface, the initial radiation emission from the explosion, termed ``shock breakout", occurs as the radiation-mediated shock \citep{weaver_structure_1976} reaches this stellar surface. When the optical depth ahead of the shock drops to $\sim c/v$,\textbf{ where $v$ is the shock velocity,} the escaping radiation produces a hours-long bright X-ray/UV flash \citep{lasher_simple_1975,klein_x-ray_1978}, followed by a days-long UV/optical emission from the expanding cooling shock-heated envelope. Existing theoretical analyses \citep[e.g.,][]{nakar_early_2010,piro_shock_2010,rabinak_early_2011,katz_non-relativistic_2012,sapir_non-relativistic_2013,sapir_uvoptical_2017,faran_early_2019,margalit_analytic_2022,morag_shock_2023,morag_shock_2024} provide a good understanding of the radiation emitted during stellar breakout and envelope cooling \citep[see][for a review]{waxman_shock_2017}.

With the advent of new wide-field high-cadence sky surveys, early light curves are accessible for numerous events, attracting significant attention and allowing us to draw initial constraints on the SN progenitors' population as a whole. As described below, the observed early light curves suggest that in many, perhaps most, core-collapse SNe, breakout occurs not at a sharp stellar surface but rather within optically thick extended material. The extended material may be a circumstellar matter (CSM) from a preexplosion mass ejection event, or, alternatively, be a low-mass extended bound envelope (that may or may not be in hydrostatic equilibrium\footnote{\citet{fuller_boil-off_2024}, e.g., recently suggested that the extended material around RSGs is a quasi-steady dense chromosphere that is supplied by outgoing shock waves, inflating the star to a few times its original radius.}). 

Various suggestions have been made for the preexplosion mass ejection origin, which is not yet well understood, including: pair instability pulsations \citep[e.g.,][]{rakavy_instabilities_1967,woosley_pulsational_2007}, binary interaction \citep[e.g.,][]{chevalier_common_2012,soker_explaining_2013}, radiation-driven instability \citep[e.g.,][]{suarez-madrigal_local_2013}, unstable late-stage nuclear burning \citep[e.g.,][]{smith_preparing_2014,woosley_remarkable_2015}, dissipation of internal gravity waves driven by core burning \citep[e.g.,][]{shiode_setting_2013,fuller_pre-supernova_2017,fuller_pre-supernova_2018}, and core magnetic activity \citep[][]{cohen_pre-supernova_2024}. Intense mass-loss episodes challenge the canonical picture \citep[e.g.,][]{langer_presupernova_2012} of a rapidly evolving core surrounded by a nearly time-independent envelope. Information on the structure of the progenitor star, on its envelope nature, and on its mass-loss history close to the explosion, is highly instructive for the study of the SN explosion mechanisms, and the impact of the progenitor structure on their different apparent types, which are not fully understood despite many years of research.

CSM breakouts are interesting both because their observations provide information on the progenitors and their preexplosion evolution, as shown recently for SN 2023ixf, and also because they may be the sources of several classes of powerful transients: they are considered as possible explanations of (at least part of) the superluminous SN class \citep[e.g.,][]{ofek_supernova_2010,chevalier_shock_2011,ginzburg_superluminous_2012,moriya_analytic_2013}, of the early emission of SNe IIn \citep[e.g.,][]{ofek_interaction-powered_2014,drout_rapidly_2014,ibik_ps1-11aop_2025}, of X-ray flashes and low-luminosity $\gamma$-ray bursts associated with SNe Ib/c \citep[e.g.,][]{tan_trans-relativistic_2001,campana_association_2006,waxman_grb_2007,soderberg_extremely_2008}, and of SNe with prominent X-ray emission lasting from days to years \citep[e.g.,][]{ofek_x-ray_2013,levan_superluminous_2013,ofek_sn_2014,chandra_circumstellar_2018}.

A recent large sample study of regular type II SNe with early optical-UV observations \citep{irani_early_2024} shows that CSM-free models account very well for about half of the population\footnote{The models constrain the radius, composition, and the explosion energy, and the derived progenitor radii are consistent with the RSG radii distribution measured locally \citep{irani_early_2024}.}, while the light curves of the other half, showing days-long rise of bright luminosity and high color temperature, are inconsistent with envelope cooling emission and suggest a shock breakout within an extended CSM at many thousands of $R_\odot$. Recently, \citet{wasserman_optical_2025} presented solutions to the problem of shock breakouts within a CSM ``wind" density profile, $\rho\propto r^{-2}$, describing the evolution of the plasma and radiation field, including the transition from radiation-mediated to collisionless shock, showing that if $R_{\rm bo}$ is not at the edge of the CSM, there is a transition from UV to X-ray emission that is not significantly suppressed by propagation through the CSM. The prevalence of dense CSM around SN II progenitors is further supported by additional growing observational evidence, such as ``flash spectroscopy", systematic analyses of precursor bursts associated with preexplosion mass ejection events, and detailed studies of the CSM structure of nearby SNe like SN 2023ixf \citep[see][and references therein for detailed observational evidence of CSM prevalence]{wasserman_optical_2025}.

Another exciting discovery is that many hydrogen-poor core-collapse SNe, sometimes referred to as Stripped-Envelope SNe (SESNe) as they are thought to have shed or been stripped of most/all of their outer hydrogen envelope, consisting of the types IIb and Ib/c, show a very early, prominent optical peak/bump, or emission ``excess" in their light curves on a day timescale. The second ``standard" peak is on a timescale of weeks and is clearly powered by the decay of $^{56}$Ni, giving these events the name ``double-peak" SNe\footnote{It is important to note that the blue bands alone can also show an early peak due to a decreasing temperature, which is generally expected for a normal massive star \citep{nakar_early_2010,rabinak_early_2011}. We therefore emphasize that we consider the events where the early peak is observed also in the red bands \citep[see Figure 1 of][]{nakar_supernovae_2014}.}. Recently, \citet{ayala_early_2025} found that $30-50\%$ of their large early detected IIb sample exhibits this early peak \citep[see also][for systematic characterization of early IIb peak properties]{crawford_peaky_2025}, and \citet{das_probing_2024} estimated it is found in almost 10\% of Ib/c.

These early peaks are usually attributed to cooling emission from shocked extended material\footnote{Alternatively, some works also explain these early peaks as a result of a double $^{56}$Ni distribution \citep[e.g.,][]{bersten_early_2013,orellana_supernova_2022,sharma_twin_2025}, an interaction with a companion star \citep[e.g.,][]{kasen_seeing_2010}, or a internal engine such as a magnetar \citep[e.g.,][]{kasen_magnetar-driven_2016}.}, with the first well-known example of the nearby double-peak type IIb SN1993J, whose progenitor was suggested to have an extended envelope of $\sim0.2M_\odot$ \citep{hoflich_sn_1993,woosley_sn_1994}. Existing theoretical analyses \citep{nakar_supernovae_2014,piro_using_2015,sapir_uvoptical_2017,piro_shock_2021}, as well as targeted numeric simulations, have been recently used to probe the mass and radius of the extended material producing these early peaks for many events \citep[e.g.,][]{bersten_type_2012,nicholl_lsq14bdq_2015,taddia_iptf15dtg_2016,arcavi_constraints_2017,fremling_ztf18aalrxas_2019,ho_sn_2020,jin_effect_2021,ertini_sn_2023,das_probing_2024,farah_shock-cooling_2025,chiba_hydrodynamic_2025,subrayan_early_2025,zou_sn_2025}. For SNe IIb, the extended material is typically assumed to be a low-mass extended bound envelope which is thought to remain around their progenitors, and may be consistent with the non-variable nature of IIb progenitors in the years before explosion, see \citet{strotjohann_search_2015} and Section \ref{subsec:SESNe}, while for SNe Ib/c, whose progenitors are suggested to be compact stars that completely lost their envelopes, the extended material is usually attributed to a preexplosion ejected CSM (although not supported by precursor evidence).

As noted also by \citet{khatami_landscape_2024}, observations with incomplete time coverage and a limited number of bands, as is commonly the case for early SN observations, can sometimes be explained equally well as, e.g., the breakout emission from extended material, and as the shock-cooling emission after significant expansion of the shocked plasma. Such different interpretations lead to widely different inferred values of the physical parameters of the progenitor and extended material. This holds for both analytic and numeric models, especially when the latter do not include a proper description of key physical processes affecting the radiation \citep[such as the generation of collisionless shocks, see discussion in][]{wasserman_optical_2025}.

While observations do not yet provide stringent constraints on CSM density structure or the ejection mechanisms, the results of numeric simulations of mass ejection following energy deposition in stars yield density distributions which are not very different from a wind profile \citep[see][and references therein]{wasserman_optical_2025}. In addition, we show that most of the emission key features are insensitive to the exact details of the density profile. We therefore analyze a simple setting of the extended material (described in Section \ref{sec:formulation}), which enables a clear analytic understanding of the different emission phases in different regimes of physical parameters, and we explain how it can be generalized for a wide range of density profiles.

The paper is organized as follows. In Section \ref{sec:formulation} we present the problem setting and parameters definitions. In Section \ref{subsec:bol_light} we derive the bolometric light curves obtained in different regimes of the parameter space, and in Section \ref{subsec:cool_T} we derive the characteristic radiation temperature during the shock cooling phase of the emission. We analyze the degeneracies in inferring the values of model parameters from observations of the bolometric light curves in Section \ref{subsec:deg_bol}, and from optical bands observations of the cooling emission in Section \ref{subsection:opt_deg}. The implications of these degeneracies for the inferred properties of the extended material around SESNe are discussed in Section \ref{subsec:SESNe}. Our results are summarized, and their implications are discussed in Section \ref{sec:conclusions}.

\section{Formulation of the Problem}\label{sec:formulation}

We consider a shock with velocity $v$ traversing a spherically symmetric material shell of total mass $M_{\rm e}$ and constant opacity $\kappa$ extending up to a sharp truncation radius $R_{\rm e}$, with a stationary wind density profile, $\rho\propto r^{-2}$.

\begin{figure}
    \centering
    \includegraphics[height=8cm,trim=1cm 0cm 0cm 0cm,clip]{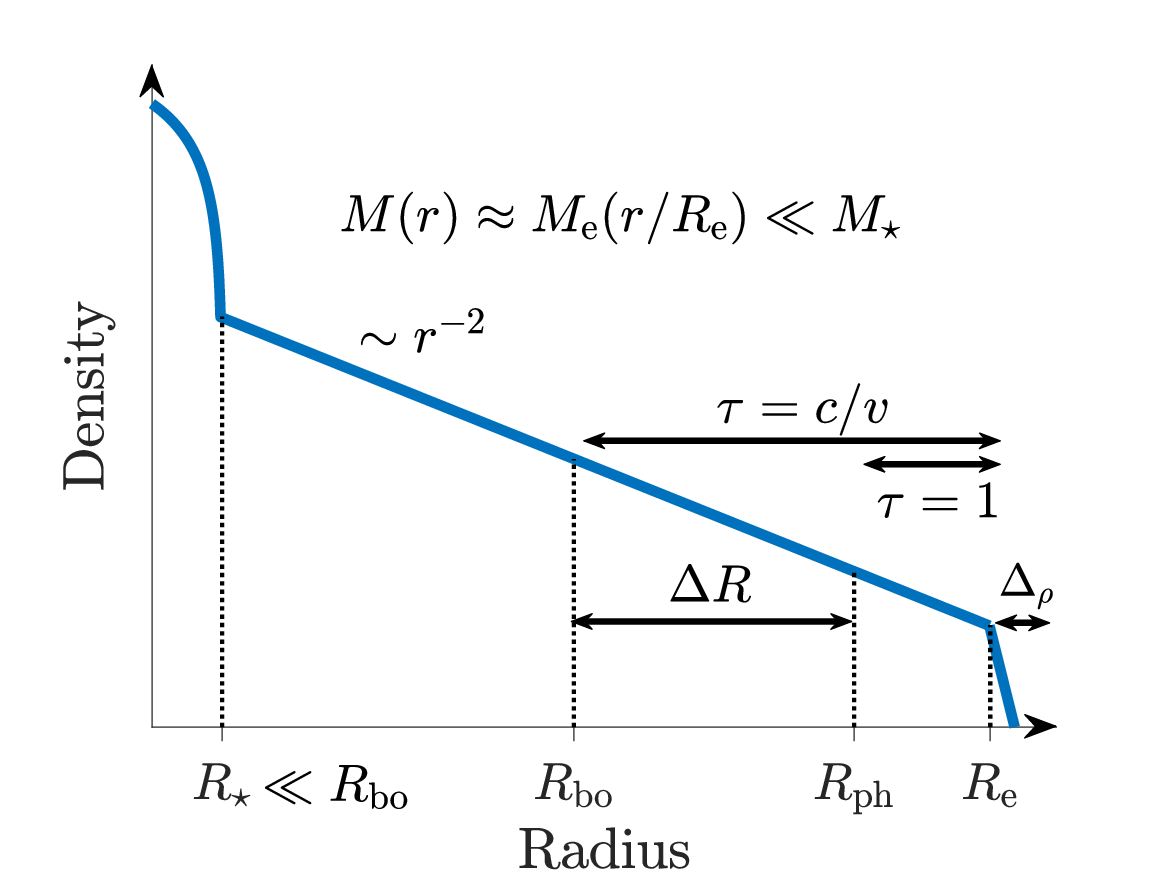}
    \newline
    \caption{A schematic description of the density profile of the extended material, with the different characteristic radii defined in the text, shown for the ``sharp truncation" case, $\Delta_\rho\ll (c/v)R_{\rm e}/\tau_{\rm e}$.}
    \label{fig:density}
\end{figure}

Figure \ref{fig:density} describes the different characteristic radii of the extended material that are defined below. 
\newline
The inner boundary of the material, which can be associated with a surface of the progenitor star (if the extended material is a preexplosion ejected CSM) or the radius of compact stellar core (if the extended material is a low-mass extended envelope) is assumed small\footnote{For an estimation of the impact on the luminosity of a larger $R_\star$ of an initial polytropic stellar profile, see the appendix of \cite{wasserman_optical_2025}.}, $R_\star \ll R_{\rm bo}$, where the breakout radius $R_{\rm bo}$ is defined below in Equation (\ref{eq:R_bo}). The extended material mass is assumed much smaller than the shock-driving ejecta mass $M_{\rm e} \ll M_\star$, such that the shock velocity is approximately constant\footnote{For analysis of the (slow) shock deceleration in the case of ejected polytropic envelope driving the shock see \cite{chevalier_self-similar_1982}, and for analysis of the strong deceleration due to a very massive CSM, as is thought to be the case in some very rare events like SN2010jl \citep{ofek_sn_2014}, see, e.g, \cite{ginzburg_superluminous_2012,svirski_optical_2012,khatami_landscape_2024}.}. In these limits, the results are independent of $R_\star$ and $M_\star$. Finally, we note that we do not include in our analysis the contribution to the light curve from $^{56}$Ni decay, and focus on the early light curve powered solely by shock interaction with the extended material.

The wind density and optical depth profiles (for $R_\star\ll R_{\rm e}$) are given by 
\begin{equation}
\label{eq:density}
\begin{split}
       \rho(r)&=\rho_{\rm e}(R_{\rm e}/r)^2,\\
       \tau(r)&=\tau_{\rm e}(R_{\rm e}/r-1),\\
\end{split} 
\end{equation}
where
\begin{equation}
\label{eq:density_val}
\begin{split}
       \rho_{\rm e}&=\frac{ M_{\rm e}}{4\pi R_{\rm e}^3}=1.6\times10^{-13}~{\rm g~cm^{-3}}\left(\frac{M_{\rm e}}{10^{-2}M_\odot}\right)\left(\frac{R_{\rm e}}{10^{3}R_\odot}\right)^{-3},\\
       \tau_{\rm e}&=\frac{\kappa M_{\rm e}}{4\pi R_{\rm e}^2}=66~\kappa_{0.2}\left(\frac{M_{\rm e}}{10^{-2}M_\odot}\right)\left(\frac{R_{\rm e}}{10^{3}R_\odot}\right)^{-2},\\
\end{split} 
\end{equation}
with $\kappa=0.2 ~{\rm cm^2~g^{-1}}~\kappa_{0.2}$. Note that $\tau_{\rm e}$ is not the total optical depth of the material, but rather the optical depth at radii $r\sim R_{\rm e}$ (for the wind density case, it is the optical depth at $r=R_{\rm e}/2$). As we show in Section \ref{subsec:bol_light}, this parameter is the most important one for shaping the light curve.

The breakout and photosphere radii, defined at optical depth of $c/v$ and unity respectively, are calculated from Equation (\ref{eq:density})\footnote{The requirement $R_\star<R_{\rm bo}$ then translates into a lower limit on the material mass $M_{\rm e}>\frac{c}{v}\left(\frac{R_{\rm e}}{R_\star}-1\right)^{-1}\frac{4\pi R_{\rm e}^2}{\kappa}$, such that for lower $M_{\rm e}$ the breakout will occur below the extended material shell.}
\begin{equation}
\label{eq:R_bo}
    \begin{split}
        R_{\rm bo}&=\left(1+\frac{c/v}{\tau_{\rm e}}\right)^{-1}R_{\rm e},\\
        R_{\rm ph}&=\left(1+\frac{1}{\tau_{\rm e}}\right)^{-1}R_{\rm e}.
    \end{split}
\end{equation}

As will be explained in Section \ref{subsec:bol_light}, we expect the resulting emission to provide the correct qualitative behavior for general density profiles where a significant fraction of the mass, $M_{\rm e}$, lies near $R_{\rm e}$, as is the case also for a polytropic profile. We note that for shallow density profiles, e.g., $\rho\propto r^0$, the optical depth does not increase at small radii much beyond $\tau_{\rm e}$, hence in this case the breakout will occur for $\tau_{\rm e}<c/v$ below the extended material shell.

The radial profile of the density in the ``truncation" region beyond $R_{\rm e}$ affects the breakout signal only if the width $\Delta_\rho$ over which the density decreases significantly (see Figure \ref{fig:density}) is large, $\Delta_\rho\gtrsim (c/v)R_{\rm e}/\tau_{\rm e}$. In this case, the breakout signal will differ from that of the ``sharp truncation" case, $\Delta_\rho\ll (c/v)R_{\rm e}/\tau_{\rm e}$.
In Section \ref{subsec:bol_light} we detail how and when deviations from the simple $\rho\propto r^{-2}$ and sharp truncation density structure affect the results, and we provide the expected modifications of the emission properties.

\section{Bolometric light curves and cooling emission temperature}\label{sec:regimes}

\subsection{Bolometric Light Curves}\label{subsec:bol_light}

As the shock velocity $v=10^9{\rm cm~s^{-1}}v_9$ is approximately independent of radius (Section \ref{sec:formulation}), the generated shock luminosity is roughly time independent\footnote{In the limit of wind breakout defined below, the shock becomes adiabatic rather than radiative, but only at very large radii $r>20\left(1+Q_{\rm brem}\right)^{1/2}R_{\rm bo}$, where $Q_{\rm brem}$ is the bremsstrahlung to Inverse Compton emissivity ratio \citep{wasserman_optical_2025}.}
\begin{equation}
\begin{split}
    L_{\rm sh}(r)&=4\pi r^2\times\frac{1}{2}\rho v^3\\
    &=1.4\times10^{44}~{\rm erg~s^{-1}}~v_{9}^3\left(\frac{M_{\rm e}}{10^{-2}M_\odot}\right)\left(\frac{R_{\rm e}}{10^{3}R_\odot}\right)^{-1},
\end{split}
\end{equation}
where we used the wind density Equations (\ref{eq:density})-(\ref{eq:density_val}). The shock crossing time of the extended material is
\begin{equation}\label{eq:t_e}
    t_{\rm e}=\frac
    {R_{\rm e}}{v}=0.8~{\rm days}~v_9^{-1}\left(\frac{R_{\rm e}}{10^{3}R_\odot}\right).
\end{equation}

\begin{figure*}
    \centering
    \includegraphics[height=7.5cm,trim=0.5cm 0cm 0 0cm,clip]{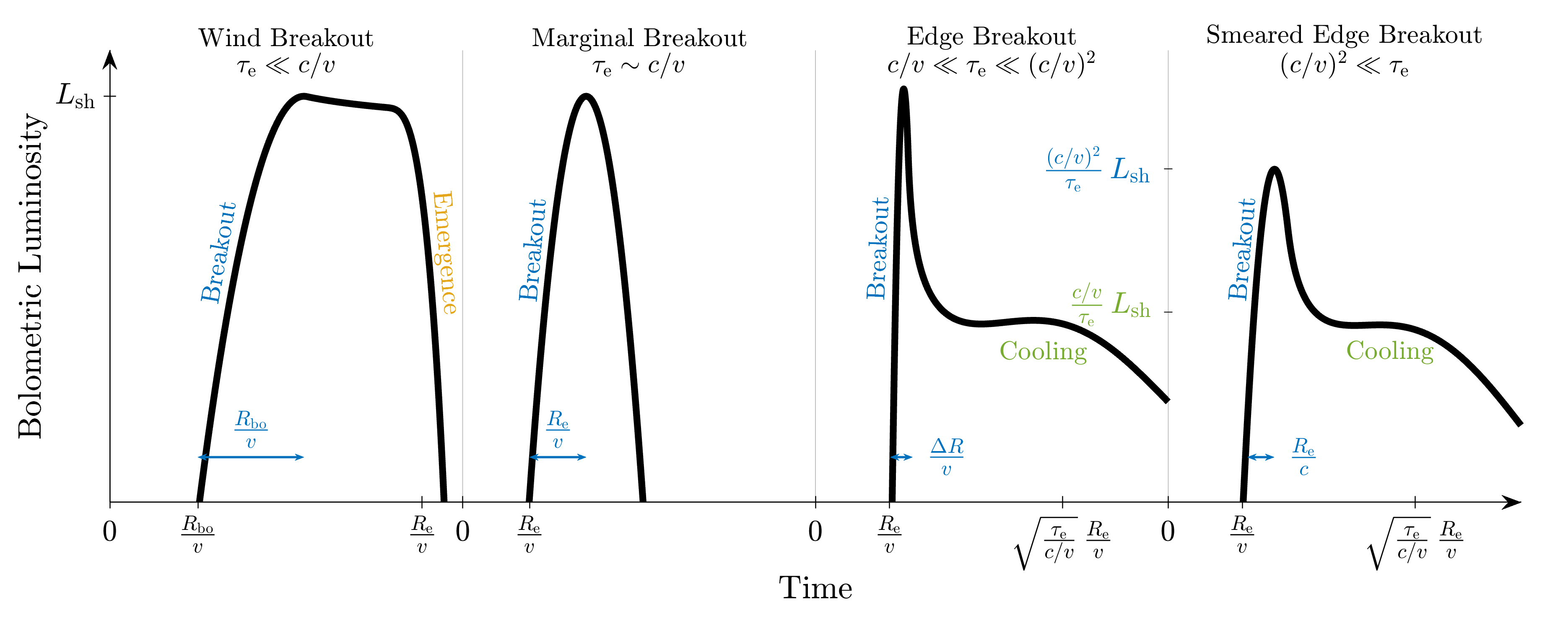}
    \caption{Schematic demonstration of early bolometric light curves powered by shock interaction with a wind extended material that is sharply truncated at $R_{\rm e}$, at different regimes of the material optical depth $\tau_{\rm e}$ that set the location of $R_{\rm bo}$ (see text). The time and luminosity scales, as well as the breakout rise time in the different regimes, are derived in the text.}
    \label{fig:lightcurves}
\end{figure*}

From the breakout radius definition, Equation (\ref{eq:R_bo}), it is evident that for $\tau_{\rm e}\ll c/v$ we have $R_{\rm bo}\ll R_{\rm e}$, while in the opposite limit both $R_{\rm bo}$ and $R_{\rm ph}$ are very close to the edge $R_{\rm e}$, with $R_{\rm ph}-R_{\rm bo}=\Delta R\ll R_{\rm bo}$. These two limits separate between two qualitatively different breakout scenarios and corresponding light curves: a UV-dominated breakout burst followed by ``cooling
emission” of the shock-heated material, and a ``wind breakout” deep inside the extended material shell with a prolonged breakout pulse followed by extended emission with spectral shift from UV to X-rays as the shock becomes collisionless while propagating up to the edge $r=R_{\rm e}$, where the shock ``emerges" out of the dense material \citep[see also][]{chevalier_shock_2011,khatami_landscape_2024}. Figure \ref{fig:lightcurves} describes the bolometric light curves obtained in these different regimes, which we now discuss in detail. 

Following \citet{ginzburg_superluminous_2012}, the average photon diffusion time from $R_{\rm bo}$ to $R_{\rm ph}$ can be estimated by considering the density-dependent mean free path
\begin{equation}\label{eq:t_bo}
\begin{split}
    \Delta t_{\rm diff}&=\int_{R_{\rm bo}}^{R_{\rm ph}}\frac{d[(r-R_{\rm bo})^2]}{c/(\rho\kappa)}=2\left(\ln\frac{R_{\rm ph}}{R_{\rm bo}}+\frac{R_{\rm bo}}{R_{\rm ph}}-1\right)\frac{\tau_{\rm e}}{c/v}\frac{R_{\rm e}}{v},\\
&=
\begin{cases} 2\left(\ln\frac{c}{v}+\frac{v}{c}-1\right)\frac{\tau_{\rm e}}{c/v}\frac{R_{\rm e}}{v}&\tau_{\rm e}\ll c/v,\\
  (1-\frac{v}{c})^2\frac{c/v}{\tau_{\rm e}}\frac{R_{\rm e}}{v} & c/v\ll \tau_{\rm e},
\end{cases}
\end{split}
\end{equation}
where we used the wind density profile, Equation (\ref{eq:density}), and the breakout and photosphere radii, Equation (\ref{eq:R_bo}), and ignored geometric corrections. This reproduces (up to order unity factors) the known limits for the breakout rise time of $\frac{\tau_{\rm e}}{c/v}\frac{R_{\rm e}}{v}\rightarrow R_{\rm bo}/v$ in the wind breakout limit and $\frac{c/v}{\tau_{\rm e}}\frac{R_{\rm e}}{v}\rightarrow (R_{\rm ph}-R_{\rm bo})/v=\Delta R/v$ in the edge breakout limit. 

\begin{figure}
    \centering
    \includegraphics[height=7cm]{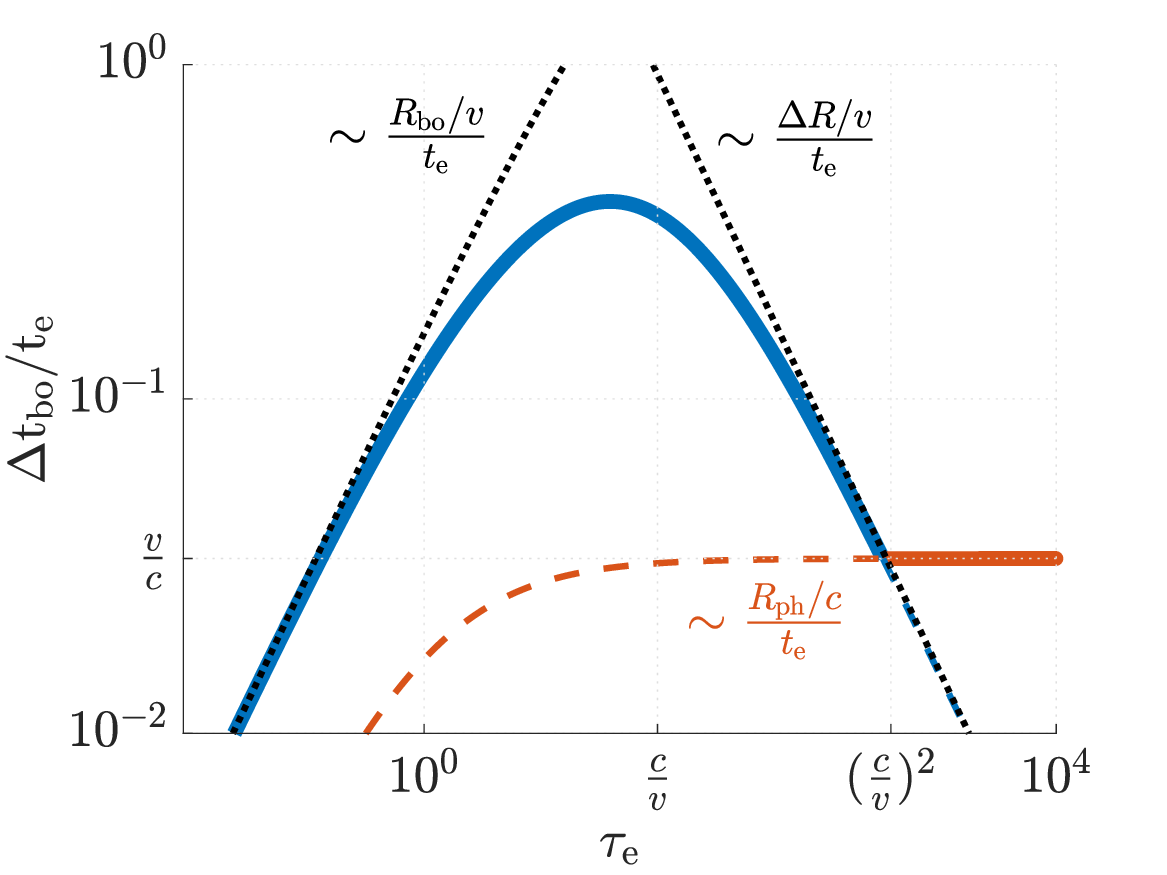}
    \caption{The breakout rise time normalized to the extended material crossing time, $t_{\rm e}=R_{\rm e}/v$, Equation (\ref{eq:t_e}), as a function of $\tau_{\rm e}$, that sets the location of $R_{\rm bo}$ (see text), with $v=10^9~{\rm cm~s^{-1}}$. For $\tau_{\rm e}<(c/v)^2$, the breakout rise time is given by the intrinsic diffusion time (blue), Equation (\ref{eq:t_bo}), with both limits of edge and wind breakouts presented in dotted black lines. For $(c/v)^2<\tau_{\rm e}$, the observed breakout rise time is smeared to the light travel time (orange).}
    \label{fig:t_bo}
\end{figure}

Figure \ref{fig:t_bo} shows the breakout rise time as a function of $\tau_{\rm e}$.
It reaches a maximal fraction of $t_{\rm e}=R_{\rm e}/v$ between the two $\tau_{\rm e}$ regimes, that can be found numerically, and for $v/c\ll 1$ it is $0.43~R_{\rm e}/v$ at $\tau_{\rm e}=0.46~c/v$. For $(c/v)^2< \tau_{\rm e}$, the observed breakout rise time is smeared by the light travel time to $R_{\rm ph}/c\approx R_{\rm e}/c$, and the observed breakout luminosity correspondingly reduces to $\frac{(c/v)^2}{\tau_{\rm e}}L_{\rm sh}$ (see right panel of Figure \ref{fig:lightcurves}, and large $\tau_{\rm e}$ in Figure \ref{fig:t_bo})\footnote{In the wind breakout limit, the photosphere location converges at $(c/v)R_{\rm bo}$ and the light travel time $R_{\rm ph}/c$ is shorter than the intrinsic diffusion time of a few $R_{\rm bo}/v$, see small $\tau_{\rm e}$ in Figure \ref{fig:t_bo}.}. We note that $\sim R_{\rm e}/c$ is the minimal light smearing time, corresponding to a perfect spherical symmetry, and deviations from such a configuration, with the shock reaching different radii at different angles, may extend the light smearing time up to $\sim R_{\rm e}/v$.

For edge breakouts in density profiles with large $\Delta_\rho$, $\Delta_\rho\gtrsim \frac{c/v}{\tau_{\rm e}}R_{\rm e}=\Delta R$, the breakout signal will be modified compared to that obtained above for the sharp truncation case, $\Delta_\rho\ll\frac{c/v}{\tau_{\rm e}}R_{\rm e}$. The breakout will take place within the material extending beyond $R_{\rm e}$, and the breakout signal properties will be determined by the density and shock velocity at the breakout position, with $\Delta t_{\rm bo}=c/\kappa\rho_{\rm bo}v_{\rm bo}^2$ and $L_{\rm bo}=2\pi R_{\rm e}^2\rho_{\rm bo}v_{\rm bo}^3$ \citep[see][who calculate the exact bolometric light curves for breakouts in polytropic envelope density profiles, showing also that they are insensitive to the exact density profile and determined mainly by the breakout density and velocity]{katz_non-relativistic_2012}. Taking, for example, a linear density profile vanishing at $r=R_{\rm e}+\Delta_\rho$, $\rho=\rho_{\rm e}(x/\Delta_\rho)$ with $x=R_{\rm e}+\Delta_\rho-r$, the breakout location $x_{\rm bo}$ is determined by $\tau(x_{\rm bo})=c/v(x_{\rm bo})$, and accounting for shock acceleration using the Sakurai self-similar solution, $v\propto\rho^{-0.19}$ \citep[][]{sakurai_problem_1960}, we find that the breakout diffusion rise time is extended compared to the sharp truncation case by a factor $2f_{\rm trunc}^{0.34}$, and that the breakout luminosity is reduced by a factor $f_{\rm trunc}^{-0.24}$, where $f_{\rm trunc}=\Delta_\rho/(2\frac{c/v}{\tau_{\rm e}}R_{\rm e})$\footnote{Note that the treatment of \citet{khatami_landscape_2024} to the case of breakout in the decaying part of the density profile beyond $R_{\rm e}$ is not self-consistent: Assuming $\rho\propto r^{-p}$ beyond $R_{\rm e}$ and taking the limit of infinite steepness $p\rightarrow\infty$, the resulting breakout signal should converge to that of the sharp truncation case as the optical depth of the material beyond $R_{\rm e}$ vanishes.}. As discussed above, the breakout pulse may instead be smeared by light travel time if the diffusion time is shorter than $R_{\rm e}/c$.

In the wind breakout regime, Equation (\ref{eq:t_bo}) can be integrated to obtain the breakout rise time for a general density profile. In this regime, the shock continues to propagate through the extended shell following breakout, producing a luminosity comparable to the breakout luminosity (tracking the density profile) until reaching $r=R_{\rm e}$, at which point the emission abruptly decays as the shock emerges from the shell (and the luminosity decay will depend on the density decay profile). During propagation, the radiation-mediated shock is converted to a collisionless shock that heats the plasma to $\sim100~$keV \citep[][]{katz_x-rays_2011}. The collisionless shock develops when the shock reaches $r\approx0.3R_{\rm bo}$, shifting most of the radiation from the UV band to X-rays by the time the shock reaches $r\gtrsim 3 R_{\rm bo}$ \citep{wasserman_optical_2025}, a radius range relevant for $\tau_{\rm e}\lesssim 0.3~c/v$ regime.

In the edge breakout regime, the breakout decay timescale is similar to the rise time. At later times, the emission is dominated by radiation escaping from the shock-heated plasma, which expands at a velocity $\sim v$. The time $t_{\rm c}$ (and radius $R_{\rm c}$), at which the emission from this expanding and cooling shell of mass $M_{\rm e}$ reaches its maximum luminosity is when the shell's optical depth drops to $c/v$,  (\citet{arnett_type_1982}, and see \citet{piro_using_2015} in the context of expanding extended material)
\begin{equation}\label{eq:t_c}
\begin{split}
    t_{\rm c}&=\frac{R_{\rm c}}{v}=\sqrt{\frac{\kappa M_{\rm e}}{4\pi c v}}=\sqrt{\frac{\tau_{\rm e}}{c/v}}\frac{R_{\rm e}}{v}\\
    &=3.8~{\rm days}~v_9^{-1/2}\kappa_{0.2}^{1/2}\left(\frac{M_{\rm e}}{10^{-1}M_\odot}\right)^{1/2}.
\end{split}
\end{equation}
This is also the timescale of emission decay after the ``cooling peak". Note that $t_{\rm c}>t_{\rm e}$ for $\tau_{\rm e}>c/v$, so that this cooling phase is indeed prominent only for the edge breakout regime (in the wind breakout regime, the extended material would not have been fully shocked by this time). The bolometric luminosity at the cooling peak is set by the initial internal energy in the extended material, and the adiabatic suppression to expansion
\begin{equation}\label{eq:L_c}
    L_{\rm c}=\frac{\frac{1}{2}M_{\rm e}v^2\frac{R_{\rm e}}{v t_{\rm c}}}{{t_{\rm c}}}=\frac{2\pi c v^2 R_{\rm e}}{\kappa}=\frac{c/v}{\tau_{\rm e}}L_{\rm sh}.
\end{equation}
It is smaller than the breakout luminosity.

For the most general density profile, there may be distinct cooling emission peaks associated with different concentrations of mass at different length scales. For example, if we consider the complete density profile of a compact stellar core surrounded by a low-mass extended envelope, it will naturally exhibit two distinct cooling peaks, corresponding to the timescales when each of the components becomes ``transparent". The most clear way to examine a general density profile in this context is to draw $dm/d(\log r)$ as a function of $\log r$, where $m$ is the enclosed mass, such that each mass concentration at a different length scale will correspond to a bump in this curve. The cooling emission properties are thus only weakly sensitive to the density profile of the extended material, as long as a mass of scale $M_{\rm e}$ is located at radii of scale $R_{\rm e}$, in other words, as long as a negligible part of $M_{\rm e}$ is concentrated at very small length scales. For power-law profiles $\rho\sim r^{-n}$, this corresponds to $n<3$ (and thus holds for the wind density), and for polytropic profiles, this holds for any polytropic index solution (of finite radius).

\subsection{Cooling Emission Temperature}\label{subsec:cool_T}

In the edge breakout regime, the bulk of the internal energy generated by the shock in the extended material evolves approximately in thermal equilibrium, such that after adiabatic suppression, the emitted radiation temperature at the cooling peak is
\begin{equation}\label{eq:T_c}
\begin{split}
    T_{\rm c}&=\left(\frac{1}{a_{\rm BB}}\frac{\frac{1}{2}M_{\rm e}v^2}{\frac{4\pi}{3}R_{\rm e}^3}\right)^{1/4}\frac{R_{\rm e}}{vt_{\rm c}}\\
    &=5.7~{\rm eV}~\kappa_{0.2}^{-1/2}\left(\frac{M_{\rm e}}{10^{-1}M_\odot}\right)^{-1/4}\left(\frac{R_{\rm e}}{10^{3}R_\odot}\right)^{1/4},
    \end{split}
\end{equation}
with $t_{\rm c}$ of Equation (\ref{eq:t_c}). This temperature justifies ignoring recombination and using a constant opacity of fully ionized material. Some works \citep[e.g.,][]{nakar_supernovae_2014,piro_using_2015,piro_shock_2021} use instead the effective temperature (at the photosphere), derived from the luminosity $L_{\rm c}$ and radius $R_{\rm c}$ of Equations (\ref{eq:t_c})-(\ref{eq:L_c}). This implicitly assumes that the radiation-matter coupling is very strong, such that photons thermalize close to the photosphere, and results in a temperature lower by a factor of $(c/v)^{1/4}\approx2$. Even in cooling emission from stellar envelopes, of much higher typical densities, the coupling is not so strong, such that the resulting observed temperature is higher than the effective temperature \citep[see][]{nakar_early_2010,rabinak_early_2011,sapir_uvoptical_2017}. This is further supported by recent numeric calculations of the temperature in these scenarios by \citet{chiba_hydrodynamic_2025}.

\section{Degeneracies in Inferred Values of Model Parameters from Observations}\label{sec:deg}

\subsection{Bolometric Light Curves}\label{subsec:deg_bol}

\begin{figure*}
    \centering
    \includegraphics[height=9cm,trim=5cm 0cm 3cm 1cm,clip]{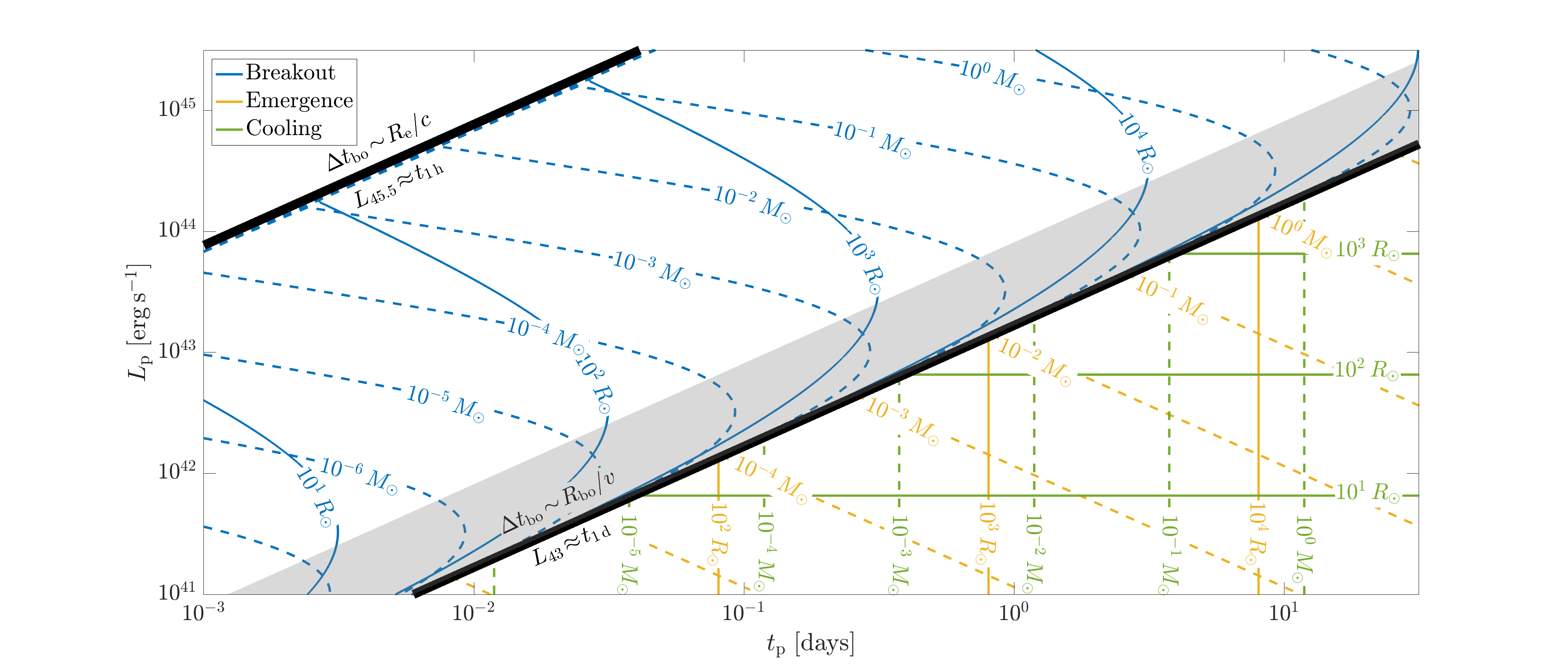}
    \caption{The bolometric peak-luminosity, $L_{\rm p}$, and peak-time, $t_{\rm p}$, obtained in the different emission phases derived in Section \ref{subsec:bol_light} as function of $M_{\rm e}$ and $R_{\rm e}$, with fixed $v=10^9~{\rm cm~s^{-1}}$ and $\kappa=0.2\,{\rm cm^2~g^{-1}}$. Contours are shown for fixed $M_{\rm e}$ with varying $R_{\rm e}$ (dashed), and for fixed $R_{\rm e}$ with varying $M_{\rm e}$ (solid).
    The lower black line corresponds to the wind breakout limit, and the Emergence and Cooling contours that cross into the (shaded) Breakout region have been removed for clarity of presentation. The upper black line corresponds to the light travel time-smeared edge breakout limit, with limiting $L_{\rm p}$ and $t_{\rm p}$ obtained by holding $R_{\rm e}$ fixed and increasing $M_{\rm e}$. The peak luminosity is limited to $L_{45.5}<t_{\rm 1h}v_9^3\kappa_{0.2}^{-1}$. Over a significant fraction of the relevant $L_{\rm p}-t_{\rm p}$ plane, the values of $M_{\rm e}$ and $R_{\rm e}$ inferred from the observed $L_{\rm p}$ and $t_{\rm p}$ depend on the assumed emission phase.}
    \label{fig:L&t}
\end{figure*}

Figure \ref{fig:L&t} shows the bolometric peak-luminosity, $L_{\rm p}$, and peak-time, $t_{\rm p}$, obtained in the different emission phases derived in Section \ref{subsec:bol_light} as function of $M_{\rm e}$ and $R_{\rm e}$, with fixed $v=10^9~{\rm cm~s^{-1}}$ and $\kappa=0.2\,{\rm cm^2~g^{-1}}$. Contours are shown in the $L_{\rm p}-t_{\rm p}$ plane for fixed $M_{\rm e}$ with varying $R_{\rm e}$, and for fixed $R_{\rm e}$ with varying $M_{\rm e}$.

Partial and/or sparse temporal coverage of the bolometric light curve, as is often the case for early SN observations (see, e.g., Figure \ref{fig:2020bvc}), may not allow one to determine which emission phase is observed, providing only a rough estimate of $L_{\rm p}$ and $t_{\rm p}$. The identification of the emission phase may be further complicated by the contribution, after a few days, of $^{56}$Ni decay. As demonstrated in Figure \ref{fig:L&t}, without an identification of the emission phase the observed $L_{\rm p}$ and $t_{\rm p}$ may correspond, over a significant fraction of the relevant $L_{\rm p}-t_{\rm p}$ plane, to different emission phases. The inferred values of $M_{\rm e}$ and $R_{\rm e}$ depend in this case on the assumed emission phase (uncertainties in $v$ and $\kappa$ further increase the uncertainty in inferred $M_{\rm e}$ and $R_{\rm e}$ values). For the longer timescales, $t_{\rm 1d}\gtrsim L_{43}$, both shock emergence and cooling phases may explain the same emission, together with the wind breakout phase in part of the region (grey shaded area in Figure \ref{fig:L&t}). For short and bright emission, $t_{\rm 1h}\approx L_{45.5}$, the light curve is controlled by light travel time and is independent of $M_{\rm e}$.

While we considered in this section the morphology of the \emph{bolometric} light curve, it should be noted that determining the \emph{bolometric} luminosity from observations over a limited wavelength range may not be straightforward, particularly since the spectrum may be far from thermal, as expected for the wind breakout regime.

\subsection{Optical Band Light Curves}\label{subsection:opt_deg}

During the cooling emission phase, the blackbody spectral peak is located near $3T_{\rm c}$, with $T_{\rm c}$ given by Equation (\ref{eq:T_c}), implying that the optical bands lie, for $M_{\rm e}<1~M_\odot\left(\frac{R_{\rm e}}{10^2~R_\odot}\right)$, at the Rayleigh-Jeans tail of the spectrum. The optical luminosity is, in this case, 
\begin{equation}
\label{eq:L_c_optic}
\begin{split}
    &\lambda L_\lambda =\frac{v}{c}\lambda L_\lambda ^{\rm BB}\approx\frac{v}{c}4\pi (v t_{\rm c})^2\frac{2\pi c}{\lambda^3}T_{\rm c}=\\
    &6.2\times 10^{41}~{\rm erg~s^{-1}}\lambda_{500}^{-3}v_9^2\kappa_{0.2}^{1/2}\left(\frac{M_{\rm e}}{10^{-1}M_\odot}\right)^{3/4}\left(\frac{R_{\rm e}}{10^{3}R_\odot}\right)^{1/4},
\end{split}
\end{equation}
where $\lambda=500~{\rm nm}~\lambda_{500}$.
While the bolometric cooling luminosity, Equation (\ref{eq:L_c}), is linear in $R_{\rm e}$ (as the radiation energy is adiabatically decreasing with expansion beyond this radius), the optical luminosity (being linear in the temperature) scales only as $\propto R_{\rm e}^{1/4}$.

The optical emission during the cooling phase is determined by the cooling time, Equation (\ref{eq:t_c}), and the peak luminosity, Equation (\ref{eq:L_c_optic}). Since the cooling time is independent of $R_{\rm e}$ and the luminosity depends only weakly on $R_{\rm e}$, the value of $R_{\rm e}$ inferred from optical observations during the cooling phase is typically highly uncertain, as small uncertainties in $v$ and in the observed $t_{\rm c}$ and $L$ lead to $1-2$ orders of magnitude uuncertainty in $R_{\rm e}$ \citep[see also, e.g.,][]{nakar_supernovae_2014,jin_effect_2021}. 

\begin{figure}
    \centering
    \includegraphics[height=5.5cm]{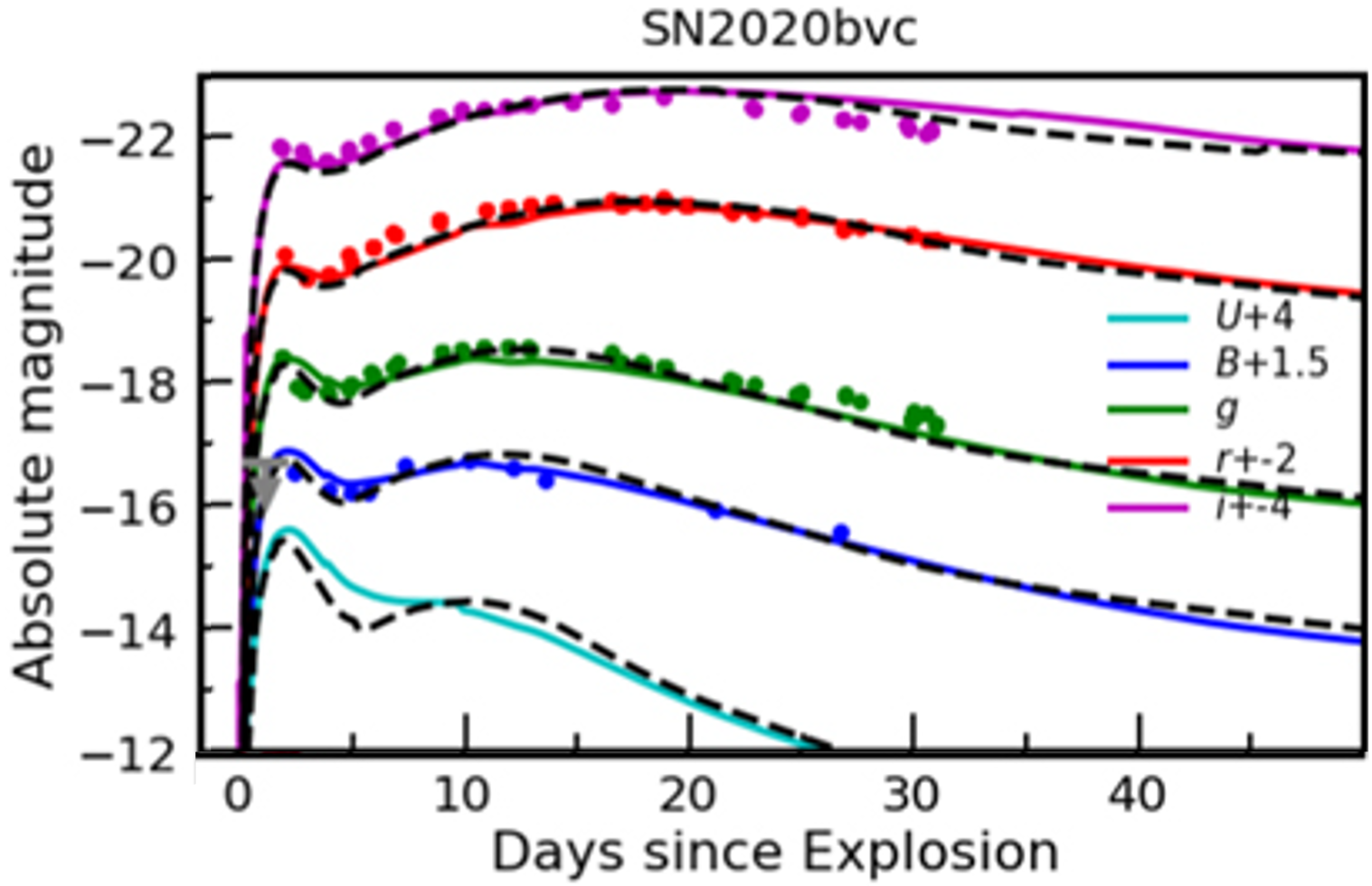}
    \caption{An example of degeneracy in the inferred values of parameters of the extended material shock cooling model: ZTF optical light curves of SN 2020bvc overlapped with STELLA radiation hydrodynamic simulations of explosions within wind density profiles (adapted from \citet{jin_effect_2021}). Solid and dashed lines represent two ``best-fit models" with $(E[10^{52}~{\rm erg/s}],M_{\rm e}[M_\odot],R_{\rm e}[{\rm cm}])=\{(1.2,0.1,10^{14}),(1.5,0.2,10^{13})\}$ respectively. The large uncertainty in the inferred value of $R_{\rm e}$ is apparent.}
    \label{fig:2020bvc}
\end{figure}

\begin{figure}
    \centering
    \includegraphics[height=8cm,trim=2.8cm 0 1cm 0,clip]{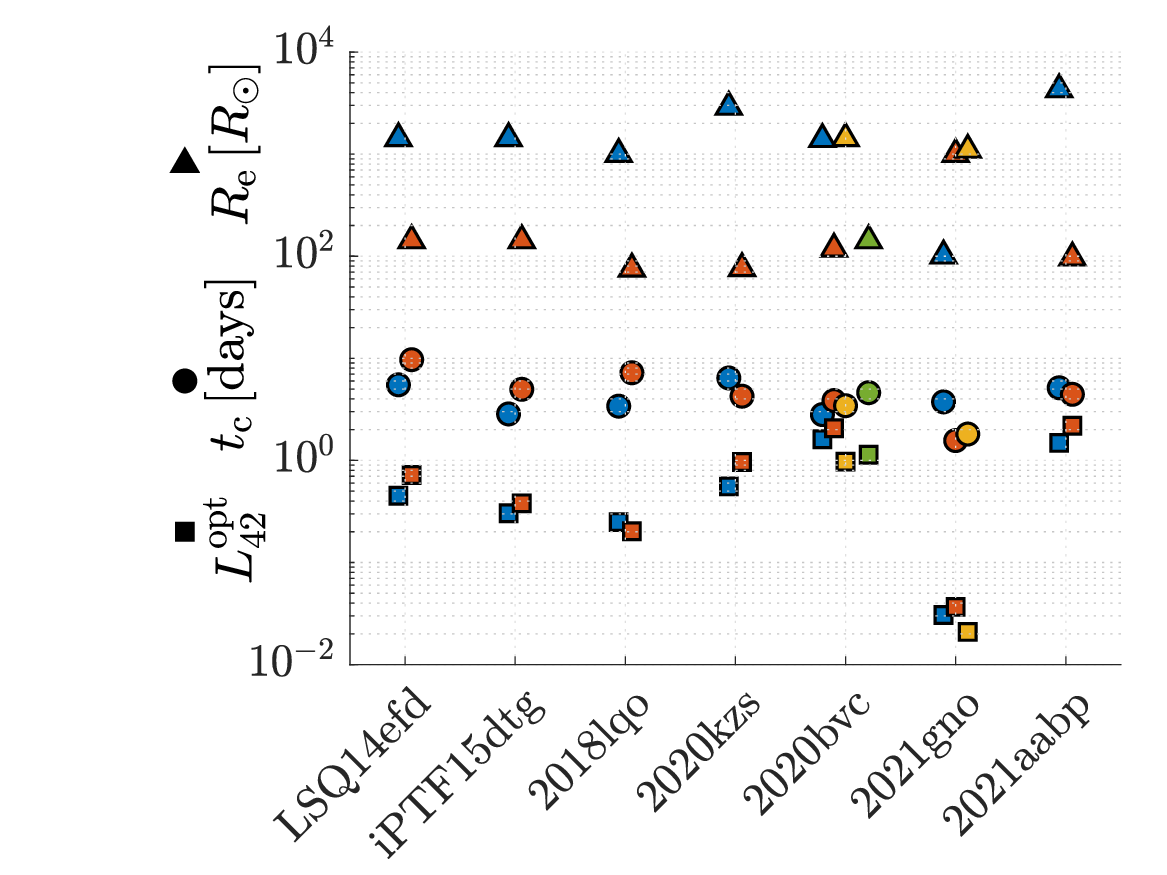}
    \caption{A selection of Stripped-Envelope SNe whose early light curve peaks have been modeled by cooling emission from shocked extended material. Different colors denote the results of different analyses, taken from \citet{jin_effect_2021,ertini_sn_2023,das_probing_2024,chiba_hydrodynamic_2025}. Optical luminosity, cooling time, and $R_{\rm e}$ values are shown in squares, circles, and triangles, respectively. The optical luminosity, $\lambda_{500}L_{\lambda_{500}}=10^{42}~{\rm erg~s^{-1}}~L_{42}^{\rm opt}$, and cooling time associated with each analysis are calculated with its $\{v,\kappa,M_{\rm e},R_{\rm e}\}$ parameter values, using Equations (\ref{eq:t_c}) and (\ref{eq:L_c_optic}). The large scatter in inferred values of $R_{\rm e}$ is apparent.}
    \label{fig:L_optical}
\end{figure}

This conclusion is demonstrated in Figure \ref{fig:2020bvc}, adapted from \citet{jin_effect_2021}. The figure shows the light curves of SN 2020bvc overlapped with their best-fit light curve, modeled by radiation hydrodynamic simulations of explosions within wind density profiles. They find in their parameter grid two sets of parameter values providing good fits, $(E[10^{52}~{\rm erg/s}],M_{\rm e}[M_\odot],R_{\rm e}[{\rm cm}])=\{(1.2,0.1,10^{14}),(1.5,0.2,10^{13})\}$. This result is roughly consistent with our analytic scalings, Equation (\ref{eq:t_c}) and (\ref{eq:L_c_optic}), that predict similar cooling times and optical luminosities for the two sets of values. This point is further demonstrated in Figure \ref{fig:L_optical}, showing the extended material shock cooling model parameter set values inferred for several SNe by different groups. The large scatter in the inferred values of $R_{\rm e}$ is evident.

\subsection{Implications for Stripped-envelope Supernovae}\label{subsec:SESNe}

In this section, we examine whether the extended material inferred to be present around SESNe based on their early light curve peaks is consistent with being an extended low-mass hydrostatic envelope.

A lower bound for the surface temperature $T_{\rm eff}$ of an extended convective stellar envelope, corresponding to an upper bound for the radius to which they inflate, is given by the Hayashi line in the Hertzsprung–Russell diagram \citep{hayashi_outer_1961},
\begin{equation}\label{eq:hayashi}
    T_{\rm eff}\gtrsim 4000~{\rm K} \iff R\lesssim 700 ~R_\odot\left(\frac{L_\star}{10^5 L_\odot}\right)^{1/2}\left(\frac{T_{\rm eff}}{4000{\rm K}}\right)^{-2}.
\end{equation}
It is determined by the temperature for which the opacity essentially ``vanishes", around $4000~$K (with a weak dependence on the stellar luminosity, mass, and metallicity), below which a sharp drop in H$^-$ opacity occurs.

The limiting value is achieved for a fully and efficiently convective star, and deviations from these conditions (which are always expected near the photosphere) tend to decrease the radius, making the above expression an upper bound. The precise value of the resulting smaller radius depends on the detailed structure of the envelope, and carries uncertainties from the mixing-length theory of inefficient convection (e.g., \citet{kippenhahn_stellar_2012} and see \citet{ou_why_2025-1} for a recent numeric analysis). Its discussion is beyond the scope of this paper.

We further note that the hydrogen deficiency of the progenitors of SNe Ib/c may modify the surface opacity and the resulting constraints for the envelope radius, although the dependence on the amount of residual hydrogen is not straightforward and requires a detailed stellar structure analysis, which we leave to future work. It is generally thought that the hydrogen (and helium) shell mass in SNe Ib/c is limited to less than $0.01-0.1M_\odot$, but it is challenging to estimate the actual amount \citep[e.g.,][]{elmhamdi_hydrogen_2006,hachinger_how_2012}.

As explained in Section \ref{subsection:opt_deg}, $R_{\rm e}$ values inferred from modeling SESNe cooling emission peaks are subject to $1-2$ orders of magnitude uncertainty (see Figure \ref{fig:L_optical}), allowing all events to be explained with $R_{\rm e}\sim10^2R_\odot$, consistent with a low-mass extended hydrostatic envelop. This is in contrast with the material inferred to extend to $>10^3R_\odot$ around progenitors of hydrogen-rich type II SNe (see Section \ref{sec:intro}), which implies a CSM resulting from a preexplosion mass ejection.  

Additional information about the structure of SN progenitors may be obtained by identifying the progenitors in existing serendipitous preexplosion images of the SN site (which is possible only for nearby, typically $\lesssim20~$Mpc events). We restrict our discussion to only ``confirmed" progenitor identifications, meaning that they include revisiting the SN site years after the SN has faded, confirming the progenitor star's disappearance \citep[see e.g.,][]{van_dyk_supernova_2017}. As of today, SESNe confirmed progenitors include four SNe IIb, a single Ib, and no Ic\footnote{Although the SNe Ib/c combined rate is estimated to be larger than the SNe IIb rate, their progenitors are thought to be optically fainter, challenging their identification \citep[e.g.,][]{yoon_nature_2012}. The small number statistics to date can also affect the number of confirmed progenitors of each type.}. The properties of these progenitors are summarized in Table \ref{tab:sn_prog}. 

\begin{deluxetable*}{l c c c c c c c}
\tablecaption{Properties of SESNe Confirmed Progenitors}\label{tab:sn_prog}
\tablehead{
\colhead{SN} &
\colhead{Type} &
\colhead{D [Mpc]} &
\colhead{$T_{\rm eff}$ [K]} &
\colhead{$\log(L_{\rm bol}/L_\odot)$} &
\colhead{$R_\star$ [$R_\odot$]} &
\colhead{Early Peak?} &
\colhead{References (and therein)}
}
\startdata
1993J        & IIb & 3.6   & $3800-4800$         & $4.8-5.4$   & $400-900$          &Yes $\sim 3~$d & \citet{richmond_ubvri_1994,maund_massive_2004}\\
2008ax       & IIb & 7.8   & $7600-20000$ & $4.4-5.3$      & $40-70$             & No ($<0.2~$d) & \citet{pastorello_type_2008,folatelli_progenitor_2015}\\
2011dh       & IIb & 7.8   & $5700-6300$         & $4.7-5.1$ & $200-300$           &Yes $\sim 3~$d & \citet{van_dyk_progenitor_2011,ergon_optical_2014}\\
2016gkg      & IIb & 18.2  & $6300-7900$               & $4.6-4.7$        & $110-160$         &Yes $\sim 2~$d & \citet{tartaglia_progenitor_2017,van_dyk_disappearances_2023}\\
iPTF13bvn   & Ib  & 22.5  & $10400-12600$    & $4.5-4.7$      & $40-60$ & No ($<0.2~$d)& \citet{cao_discovery_2013,eldridge_disappearance_2016}\\
\enddata
\end{deluxetable*}

While conclusions from such a small number of events are still not robust, and the estimated progenitor properties may suffer from large uncertainties (mainly from distance, extinction, and light blending uncertainties, see in the references of Table \ref{tab:sn_prog}), the observations appear to be consistent with the expectation that an early, day timescale, cooling peak will be observed for giant progenitors with hydrostatic envelopes extending to  $\sim10^2R_\odot$, and will not be observed for more compact envelopes. This was suggested for SNe IIb, with \citet{chevalier_type_2010} designation of cIIb and eIIb for ``compact" and ``extended" progenitors, with a dividing envelope mass of $\sim0.1M_\odot$. We propose that at least some of the Ib/c SNe that have a day-scale peak are also a result of some remaining extended hydrostatic envelope that has not been fully stripped, making them a bit closer to the IIb type. This may be a natural consequence of the canonical picture of continuity between II-P$\rightarrow$II-L$\rightarrow$eIIb$\rightarrow$cIIb$\rightarrow$Ib$\rightarrow$Ic SNe types \citep[e.g.,][]{nomoto_evolution_1995}, governed primarily by the remaining envelope of the progenitor at the time of explosion. This may also be consistent with the estimated rate at which these types show a day-scale peak (which seems less frequent for Ib/c relative to IIb, see Section \ref{sec:intro}), but further rate analysis is required.

If the extended material is the result of a preexplosion mass ejection, a precursor burst associated with the mass ejection would be expected. Such preexplosion bursts were observed in systematic analyses of precursor emission in a large sample of SNe of type IIn, finding that significant precursor emission is common during the $\sim90$ days preceding the SN explosion \citep{ofek_precursors_2014,strotjohann_bright_2021}. In contrast, no significant precursor emission is observed in systematic analyses of SNe IIb \citep{strotjohann_search_2015}. Even if the precursor burst is not observed, one may expect a strong evolution of the progenitor appearance. Out of the five identified progenitors presented in Table \ref{tab:sn_prog}, multiple-epoch preexplosion images were obtained only for 1993J, 2008ax, and 2011dh, which show no significant evolution or variability.

\section{Conclusions \& Discussion} \label{sec:conclusions}

In this paper, we showed that the bolometric light curves resulting from explosions within an extended material may be divided into two classes: edge breakouts followed by cooling emission from the expanding material obtained for $\tau_\mathrm{e}\gg c/v$, and wind breakouts followed by a continued interaction of the shock up to the extended material edge obtained for $\tau_\mathrm{e}\lesssim c/v$. These two regimes naturally link between CSM breakouts that are thought to be ubiquitous for hydrogen-rich core-collapse SNe, and cooling emission peaks that are frequent in SESNe (see Section \ref{sec:intro}). The dependence on $\{v,\kappa,M_\mathrm{e},R_\mathrm{e}\}$ of the duration and luminosity characterizing the different phases of emission were derived, see Equations (\ref{eq:t_bo})-(\ref{eq:L_c}) and Figures \ref{fig:lightcurves}-\ref{fig:t_bo}. 

We demonstrated (Figure \ref{fig:L&t}) that observed early light curve peaks/bumps may be attributed to different phases of the emission, implying a degeneracy in the inferred values of model parameters. The projection from the bolometric light curve into different wavelength bands is crucial but is not always straightforward, as the emitted radiation spectrum may be far from thermal. This is particularly important in the wind breakout regime, where the collisionless shock may drive most of the radiation energy into the X-ray band.

We derived the radiation temperature in the cooling emission phase, Equation (\ref{eq:T_c}), and showed that when observing this phase with only the optical bands, there is a $1-2$ orders of magnitude uncertainty in the inferred extended material radius $R_{\rm e}$, see Figures \ref{fig:2020bvc}-\ref{fig:L_optical}. Thus, while some works argue that large radii of many $10^3R_\odot$ are required for explaining the early emission of SNe Ib/c \citep[e.g.,][]{chiba_hydrodynamic_2025}, we showed that optical data alone cannot exclude much smaller radii of order $10^2R_\odot$. For these non-extreme radial extents, the extended material around SESNe may be a remaining extended low-mass bound envelope rather than a CSM resulting from preexplosion mass ejection. This idea seems consistent with the (very few) confirmed identifications of SESNe progenitors in preexplosion images, see Table \ref{tab:sn_prog} and Section \ref{subsec:SESNe}.

Early SN detections, with early multi-band follow-ups, as well as analyses of the theoretically expected radiation spectrum in each regime and phase of emission, are necessary for obtaining observational constraints on the progenitor and extended material parameters, which will allow a complete mapping of the properties of the SN progenitor population. In particular, for early cooling emission peaks, as the optical bands capture only a fraction of the luminosity and most radiation emerges in the UV, \emph{ULTRASAT} \citep[][]{shvartzvald_ultrasat_2024} will allow both early detection and direct measurement of the early high color temperature, Equation (\ref{eq:T_c}), lifting the degeneracy and constraining the radial extents and the nature of the extended material producing these peaks.

Additional information on the structure and origin of the extended material can potentially be obtained from ``flash spectroscopy", i.e., from early-time spectral measurements \citep{gal-yam_wolfrayet-like_2014}, which often reveal the presence of strong, narrow spectral lines of highly ionized species that disappear within a few days after explosion \citep{gal-yam_wolfrayet-like_2014,khazov_flash_2016, yaron_confined_2017, zhang_sn_2020,bruch_large_2021,terreran_early_2022,jacobson-galan_final_2024}. Such lines are most naturally explained as originating in a compact, low-velocity CSM shell that is ionized by the early UV/X-ray emission and subsequently swept up by the SN shock within a few days \citep{yaron_confined_2017,dessart_explosion_2017,boian_diversity_2019}. \cite{bruch_prevalence_2023} find that $>50\%$ of Type II SNe likely show these CSM features.

The presence, shape, and temporal evolution of these early spectral features depend on the structure of the dense CSM and on the ionizing radiation generated by the shock (and at later times on the emission from the expanding, cooling stellar envelope). The ionizing radiation, in turn, depends on the CSM structure and evolves strongly in time due to the RMS-collisionless transition, as demonstrated in this work. SN 2023ixf spectra, for example, show an increase in ionization level during the first days \citep{zimmerman_complex_2024}, consistent with a hardening of the shock-generated ionizing radiation. The results presented in this work enable a self-consistent calculation of the shock-generated continuum radiation and the resulting time-dependent spectral features, which may facilitate breaking the degeneracies between different emission regimes and model parameters.

\begin{acknowledgments}
We thank Boaz Katz and Or Guttman for insightful comments. This research was partially supported by ISF and Rosa and Emilio Segre Research Awards.
\end{acknowledgments}

\bibliography{references}

\end{document}